\documentclass{report}
\usepackage{graphicx}
\usepackage{amssymb}
\begin{document}

\begin{titlepage}
\begin{center}
{\LARGE{Blank measurement based\\ time-alignment in LC-MS}\\}
\date{}
\vspace{13mm}
Jan Urban\\
\vspace{13mm}



Laboratory of applied system biology, School of Complex Systems (former Institute of Physical Biology), South Bohemian Research Center of Aquaculture and Biodiversity of Hydrocenoses, Faculty of Fisheries and Protection of Waters, University of South Bohemia in \v{C}esk\'{e} Bud\v{e}jovice\\ Z\'{a}mek 136, Nov\'{e} Hrady 37333, Czech Republic.\newline\\

\end{center}
\end{titlepage}

\newpage
\section*{Acknowledgment}
\label{acknowledgment}
\addcontentsline{toc}{chapter}{Acknowledgment}


This work was supported and co-financed by the South Bohemian Research Center of Aquaculture and Biodiversity of Hydrocenoses (CENAKVA CZ1.05/2.1.00/01.0024); by the South Bohemia University grant GA JU 152/2010/Z; by the Ministry of Education, Youth and Sports of the Czech Republic under the grant MSM 6007665808; by the ERDF and by the INTERREG IVC programme,  project Innovation 4 Welfare, subproject PICKFIBER.

\newpage
\section*{Abstract}
\addcontentsline{toc}{chapter}{Abstract}
Here are presenting the blank based time-alignment (BBTA) as a strong analytical approach for treatment of non-linear shift in time occurring in HPLC-MS data. Need of such tool in recent large dataset produced by analytical chemistry and so-called omics studies is evident. Proposed approach is based on measurement and comparison of blank and analyzed sample evident features. In the first step of BBTA procedure, the number of compounds is reduced by max-to-mean ratio thresholding, which extensively reduce the computational time. Simple thresholding is followed by selection of time markers defined from blank inflex points which are then used for the transformation function, polynomial of second degree, in the example. BBTA approach was compared on real HPLC-MS measurement with Correlation Optimized Warping (COW) method. It was proved to have distinctively shorter computational time as well as lower level of mathematical presumptions. The BBTA is computationally much easier, quicker (more then 1000$\times$) and accurate in comparison with warping. Moreover, markers selection works efficiently without any peak detection. It is sufficient to analyze only baseline contribution in the analyte measurement with sparse knowledge of blank behavior. Finally, BBTA does not required usage of extra internal standards and due to its simplicity it has a potential to be widespread tool in HPLC-MS data treatment.

It is described in details, mathematically and experimentally justify approach for time alignment of LC-MS spectra using blank measurement data as (inherent) internal standards (BBTA). BBTA utilizes solvent contaminants and other important events (inflex points) detectable both in blank run and the compared experiment for alignment of multiple 2D chromatograms. Addition of internal standards may increase number of data points available for calculation but is not necessary for general laboratory practice. Obvious advantage of BBTA is its readiness and essentially low expenditure level of its application. All mathematical descriptions are derived immediately from the system based description of the measurement data sets with respect to the common used definitions.

%


\newpage

\renewcommand{\chaptername}{Contents}
\label{sec:contents}

\addcontentsline{toc}{chapter}{Contents}
\tableofcontents

\newpage
\section*{Introduction}
\addcontentsline{toc}{chapter}{Introduction}
The comprehensive comparison of complex mixtures of similar compounds by HPLC-MS has been major issue in 1980s and 1990s (\cite{Hearn, Snyder1, Snyder2, Mant}) and became again highly interesting with extension of so-called -omics approach from genomics to proteomics and metabolomics. There, LC-MS is one of the prime experimental tools. In this work, it is focused on measurements time alignment for comparison of multiple compounds in similar samples. For that, it is used the markers from selected spectra and the retention time values. 

In many cases of complex samples, it is recognized as crucial, difficult and nontrivial task to compare two or more measurements obtained by LC-MS. Even the measurements of samples identical in content but differing in amounts of applied quantity on the same chromatographic column with the same experiment settings are affected by nonlinear shifts in retention times. Therefore, the 'same' results do not fit together in the time axis and to compare samples, it is required transformation (normalization) function(s) to compare retention time values and other characteristics. Because of nonlinearity of the shift(s), also the normalization function has to be nonlinear. 

Naturally, the liquid phase interaction during the analyte measurement are sample dependent. Therefore, issues of those interactions are not necessarily represented in the blank. However, the processing is based on the opposite point of view. The compounds, presented in the blank are also still presented in the analyte measurement. The basis for this are trivial. Semi-similar samples (like in metabolomics) or concentration curves require sequences of analysis with the same settings, especially baseline contribution. Therefore, pertinent features pinpointed from the blank remains in the analyte measurements. They are, usually hidden in the noise contribution or peaks behavior in Total Ion Chromatogram (TIC), which is just the summary projection in one axis and therefore mathematically loss operation. However, in 3D data matrix space are still observable and detectable. Concisely, what is in the blank have to also be in the analyte measurement when the same liquid phase is used, out of the question. There should be also some shift of the shifts of the retention time values for certain elution  according to the temperature. Small changes affect only the distance of the shifts, not the ordering and it is strictly recommended to keep the conditions constant for repetitive experiments. Therefore, temperature changes in comparable measurements are also similar from the principality (and occurred in corresponding parts of the measurement). Theoretically, ordering transpositions in retention time will be caused by the huge temperature changes between the samples. Thus, the presumption of samples similarity is hardly fulfilled and it is not beyond the scope of this work. Therefore, one can simply assume that the temperature attribute is not important for the time alignment.

When corresponding retention time values are available, it may compared the peak positions by so-called Dynamic Time Warping (DTW). This is a class of signal processing method to measure similarity and find optimal match between time axes. Warps produce highly reliable output across the different measurements. Namely, when the dataset is dominated by highly similar compounds (i. e. standards). The algorithms have heavy computational burden. DTW is based on re-calculation of main part of the original dataset. Crucial aspects of warps are discussed in details later.
However, in empty (or blank) run some relevant (inflex and marker) data points may be identified (not necessarily the peaks). Blank in the context of this work is the chromatographic measurement without addition of the sample. So, it is usually just the mixture of solvents, sometimes called baseline, mobile phase or systemic noise. Hence, the blank is easily obtained for every kind of experiment and is often recorded without any utilization for experiment evaluation. Such typical data points from blank are also present in datasets from real sample analysis performed technically under the same conditions. Instead of both DTW and IS, information from the blank measurement is available for simple and immediate comparison of samples.  

The key idea of the approach presented in this work is following: The common view of the LC-MS data considers that mobile phase complicates (negatively affects) the analysis of the measurement. It contributes to random noise and it is major cause of the systemic noise (ridges and interfering peaks) in nonlinear level on the time axis. Several works are focused on removal of baseline presence from the measured data (\cite{Urban, Johnson}). The blank measurement can be considered as a permanent standard. The blank time axis has direct relation (homomorphism in fact) to all of the samples measurements obtained with the same settings, the same devices and the same mobile phase. Moreover, rapidly lower amount of relevant data points is needed to enter the computation process. Simply, one had an inherent set of internal standards. 

This work is focused on the study of the key idea to use the data from blank measurement directly for time-alignment, without any peak detection. It is done prior to any further and superfluous analysis and is of general character. The application of internal standards (IS) only adds additional information to it (mathematically just increase the amount of inflex points in the measurement). It is demonstrated on example that blank based approach is very robust, when only few presumptions are fulfilled. 

\newpage
\section*{Motivation}
\addcontentsline{toc}{chapter}{Motivation}
Liquid chromatography (LC) in tandem with Mass spectrometry (MS) is widely used in many chemical and biochemical analytical setups, especially in  so-called omics science to analyze the content of measured samples (\cite{McMaster, Ardrey}). Systems biology is important field of biological science focused on the individual components at each level of the living organisms (\cite{Noble}). The omics technologies make the systems biology realistic and experiment based science. They reveal hidden properties of the compounds present in the biological samples. Metabolomics, proteomics or lipidomics lies in the heart of gene products profile identification. LC-MS measurement is one of the key tool for the biochemical pathways analysis (\cite{Weckwerth}). 

The compounds of interest (analytes) are found as complex mixture in the sample an LC decrease the complexity by improving analyte separation. That produce the time element of the measurement, called retention time (RT). Separation process shows shifts and distortions in the RT when two or more measurements are compared. This fact makes the assignment of similar compounds difficult, since the mapping to each other is not known in advance. But it is crucial to correct for those warps. Otherwise, it is hard or even impossible to find the corresponding partners (\cite{Lange}).

Current philosophies for time normalization are divided into two major categories: Statistical models (MVA, DTW, Peak detection) and empirical rules based on internal standards. Actually, there is no restriction for the model to be based on internal standards (IS). Recently, there were developed methods for estimation of semi-optimal set of single or multiple IS, like NOMIS (\cite{Sysi-Aho}) or excellent idea of Linear solvation energy relationships (LSERs (\cite{LiJ})). The LSERs is based on selection of open windows in the chromatograms for prediction of IS candidates. This is time (and standards) saving approach which minimize the errors of samples and IS compounds mutual influence or competitions. However, both ways (NOMIS and LSERs) demands to think about it before the own measurements. Also a few of forgone experiments to choose the proper set of standards for given samples, column or method(s) are required. Let just remind that measurement improvement by standards slightly increases the total amount of required scientific budget as well as spent of time in lab.

On the other hand, the non-supervised models and derived algorithms are based on time warping approaches (\cite{Lange}). It all started with the Dynamic time warping (DTW) in speech recognition tasks. The main idea is on partial shrinking and stretching of the time axis. Naturally, reference set or piecewise transformation differ in several warping techniques. Namely, the parameters for the transformation function are in Linear time warping, Fast dynamic time warping, Parametric Time Warping (PTW) and Correlation Optimized Warping (COW) determined by maximizing or minimizing the sum of coefficients between data segments in pairs of samples (\cite{Lange, Tomasi, Chae, Hoffman, Salvador}). Time warping algorithms separate the time dimension into segments but preserve the temporal order.

Soon or later, the segmentation task leads to the peak detection problem. Strong peaks candidates allows the alignment additional flexibility (\cite{Hoffman, Norton}). Robust peak detectors require advanced analysis like noise filtration, baseline subtraction, pattern recognition or curve fitting (\cite{Norton, Johnson, Chae, Urban}). However any error in peak detection is propagated into time alignment. While this methods are effective for simple samples, they could be insufficient for more complex biological analytes (\cite{Norton}).

Nowadays extremely modish approach is principal Multivariate data analysis (MVA), especially its Principal component analysis (PCA) (\cite{Martens1, Martens2, Johnson}). It is a method of classification based on correlation and linear combination. It finds a new coordinate system from the original variables. PCA advantages are mainly the reduction of dimensionality of the data sets and better visualization of major trends in the data. It has to be realized, that two principal components were comparable only if they represent exactly the same linear combination. That is hardly fulfilled in completely different mass spectra (with possible exception only for noise presence). However, PCA is powerful mathematical tool when it is used with wisdom. 

For completeness sake, exhaustive survey of possible alignment approaches was done by (\cite{Kirchner} and \cite{Podwojski}). Recently, was published (\cite{Urban}) an information-based approach for extraction of spectra of LC-MS data, which reliable detect peaks, random and systematic noise (ridges) and store them and their statistical properties. Apart from electrical spikes, the whole spectra may be reconstructed from resulting dataset without loss of existent information. Certainly it rely on accepted model of LC-MS process, but it already introduced many amendments to it which can only make the model compatible with available data. For the first step in the whole analysis, the retention time alignment have been developed a method which is completely model-independent. This comparison is naturally more comprehensive than IS and does not require any compound identification. In some aspects, namely when abundant peaks are present, it preserve reliability. And it is shown that it is more robust than any method known to us.

\newpage
\section*{Approach}
\addcontentsline{toc}{chapter}{Approach}
Seemingly, correct and preferable approach is to use of internal standards (IS), i.e. the addition of known substance(s) into the sample(s) (\cite{Ardrey, Sysi-Aho, LiJ}). At the best, these samples should be isotopically labeled versions of the same compound. This approach may become extremely expensive, time and experimentally demanding. Often, the design of standards follows certain logic, i.e. hydrophobicity index (\cite{Krokhin}). There is no universal set of standards which would map the behavior of any solvent mixture on any column. As well as, there is no idealized column which would separate compounds only according to one chemical parameter, often also idealized. Also dynamic parameters, rate of binding of a compound to the column and release from it and column capacity affect the retention time of all compounds which interact with the column at a time. From this point of view, some combination of standard compounds may even be misleading. In practice, IS are much less often applied than they would be needed. In some cases, they are not applicable due to lack of adequate standards. 

Addition of known substance to the measured sample relieves to quality of measurement (\cite{GoldBook}). However, the addition itself is not obviously easy, exact substances selection depends on the current measurement (\cite{Bijlsma}). It has to differ from analyte, which could be a priori unknown in study of chemical fingerprints of specific processes like matabolite profiles (\cite{Daviss}). 
Nevertheless, obtained data output still require computation to fit internal standards response from slightly different measurements together. This step can not be skipped and the addition helps only (but substantially) to locate the marker data points or statistical parameters (\cite{Sysi-Aho}) for the retention time alignment.

With this knowledge and without any other assumption one can put the following question: Where to look for internal standards fulfilling the condition to be 'friendly' (different, detectable, known properties, etc.) to given sample and experiment method? The most simple answer is usually neglected for no reason. Obviously, the baseline consist of substances with very relevant features: designated amount (rate, gradient) of solvents, known or predictable affection to the analyte(s), pertinence to the column, and therefore to the requested chemical separability and specific time of elution above all.

Mobile phase in LC-MS negatively affect the measurement analysis, represent the systematic noise in nonlinear level on the time axis. However, the omitting presence of the baseline can be turn into the  advantage considering it as the permanent standard addition measurable also alone in the blank. Therefore, it worths for considering at the beginning of rough development of semi-optimal sets of internal standards or advanced comparison algorithms. Hence, the blank is easily obtained for every kind of experiment and is often done without any further use.

\newpage
\section*{Methods}
\addcontentsline{toc}{chapter}{Methods}

The reason, why the set of internal standards present in blank LC-MS measurement is so extensive, comes from measurement practice. The sample with solvent mixture is injected into a chromatographic column in LC-MS for the first separation and, due to the interaction with the column stationary phase, elutes at different retention times (\cite{Ardrey, Lange}). It is strictly recommended, but not always followed, to wash-out (clean up) the column for re-equilibration at the end of the measurements. The true wash-out takes as much as 24 hours (\cite{McMaster}), for this reason there are done only partial (short-time) wash-outs to remove the solvents and other impurities (rests of the sample, phthalate esters from preparation plastic dishes, etc.) at the end of every measurement. 
Therefore, one obtain in most measurements at least one of these events, solvent (injection) peak (SP) and/or wash-out tail (WOT). If these part(s) of data were recorded, it is another question, let assume the solicitous operator. It can not save the time of measurement to despise the beginning or end of the data. It is already done, so there is no reason for uncollect it.
In the blank measurement is SP or WOT (or both, in optimal case) the semi-dominant part of chromatogram, even if the number of solvents in mixture is small. And, because of usage of the same settings, SP or WOT has to be also presented in the sample(s) measurement, perhaps less distinguishable. In given experiment series, due to incomplete wash-out of the column, some of the solvent contaminants may (and do) actually arise from samples (or blank) themselves. Thus, their use as effective internal standards is obvious.

In this way, the time axis of the blank measurement is considered as reference time axis. It is congruent for all other sample measurements, which are done using the same settings and devices. The time-alignment consist of three main steps, each of them can be investigated by many different methods (already existed or developed in the future). It is shown here a simple but efficient example to prove the usability of the blank data, that is the key idea.

In this chapter, it extend in details all steps. All relevant issues are precisely and mathematically described and justified.

\begin{figure*}[ht]
\centering
\includegraphics[scale=0.33]{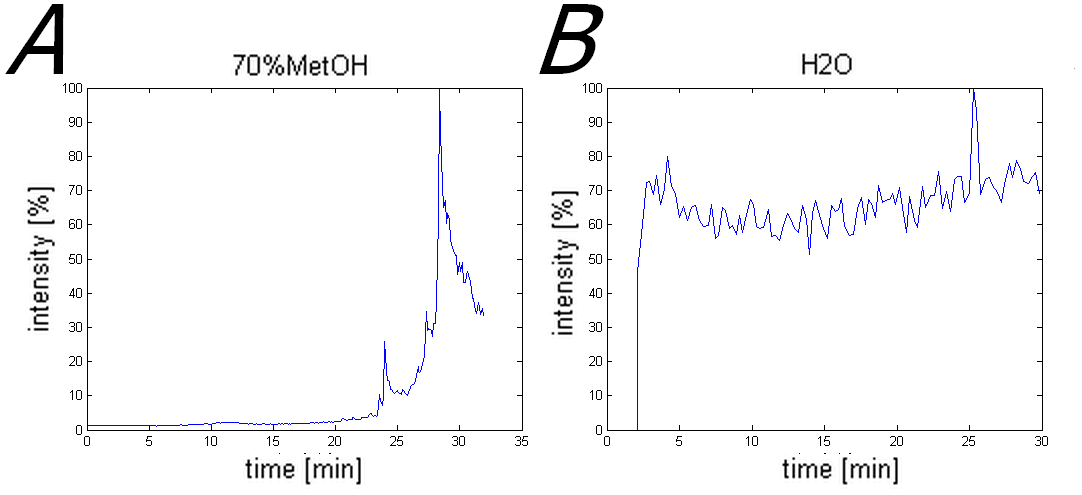}
\caption{Two examples of blank measurements. Panel 1A shows the 70\%MetOH mobile phase without solvent peak and with wash-out, panel 1B shows the H20 mobile phase with solvent peak far from ideality and without wash-out.} 
\label{blanks.graphicx}
\end{figure*}

\subsection*{Step 1.:Reduction of blank points}
The blank measurement as well as any LC-MS measurement (considering without $msn$ or other extensions) produce data of three discrete axes: retention time, mass-to-charge ratio and intensity. In other words, one obtain one intensity for each time and mass pair. This could be mathematically described as mapping from the set $T$ of time values $t$ and set $M$ of mass values $m$ into set $Y$ of intensity values $y(t,m)$. It is more transparent when the sets $T$, $M$ and $Y$ are ordered, in the following text is considered that property and all sets are ordered increasingly.
The LC-MS measurement is therefore defined by the sets $(T, M,Y)$. Let mark the sets, that defined the blank measurement as $(T_B, M_B, Y_B)$ to distinguish them in the following text from the experiment (analyte) measurements $(T_{A1}, M_{A1}, Y_{A1})$, $(T_{A2}, M_{A2}, Y_{A2})$ and so forth.

In the very first step, it is helpful to decrease the number of mass values in the blank. The reason is obvious, even the blank measurement is affected by the random noise and mass spikes. Only the true mobile phase compounds are required for the following computation. Furthermore, it is not a big pay to lose very small (in amount) compounds. They are probably just impurities, may not be present in the real sample measurement(s) and contribute in useless increase of the computation time.

The basic way, how to reduce amount of blank data points is to discard all intensity values under some thresholds value. This threshold could be general for whole blank or adaptive (different thresholds for different regions of blank), based only on the intensity value or computed via statistical parameters (PDF estimation, between-class variance, MVA) and other advanced techniques (entropy, space transformations, morphological segmentation).
For the used purpose, to show the usability of blank measurement for time alignment, is enough to compute general threshold from statistical moments. Actually, the precision of this step is not as important as in the next two steps. Decrease of data points for marker selection is more significant for computer memory (which limitation could be overcome by HDD swapping) then for the total time of computation, using todays CPUs and/or GPUs.

Let analyze individual mass $m_b \in M_B$ in the time axis and compute the maximal intensity value ${\cal X}_Y$:
\begin{equation} 
{\cal X}_Y(m_b) = max( y(t,m_b)),~t \in T_B, ~y \in Y_B
\end{equation}

and mean intensity value $\mu_Y$:
\begin{equation} 
\mu_Y(m_b) = mean( y(t,m_b)),~t \in T_B, ~y \in Y_B. 
\end{equation}

As an input for thresholding process is used max-to-mean ratio $R$ as standard method for automated data processing and observation (\cite{Chang, Bunting, Kohl}):
\begin{equation} 
R(m_b) = {\cal X}_Y(m_b) / \mu_Y(m_b). 
\end{equation}

Now, are computed two numbers from the max-to-mean ratio $R$ (with a priori unknown distribution) using statistical moments.
The number that separating the lower half of a sample from the higher half is the median, mathematically the value $\alpha$ that minimize
\begin{equation}
E(\left|R(m)-\alpha\right|),
\end{equation}
where function $E(\xi)$ is considered as the average of its argument $\xi$ (and in this case is $\xi = \left|R(m)-\alpha\right|$). Therefore, median $\alpha_{med}$ id defined as
\begin{equation}
\alpha_{med}~:~ E(\left|R(m)-\alpha_{med}\right|) = min, ~\forall \alpha_{} \in \mathbb{R},
\end{equation}
where $\mathbb{R}$ is set of real numbers.
As a measure of the variability is used robust standard deviation ($RSTD$), because the max-to-mean ratio $R$ has a priori unknown distribution:
\begin{equation}
RSTD = 1.25 * E(\left|R(m)-\alpha_{med}\right|).
\end{equation}
The threshold value $\Theta$ for max-to-mean ratio $R$ is set as
\begin{equation}
\Theta = \alpha_{med} - RSTD. \label{eq:th}
\end{equation}
Consequently, all masses $m_b$ with ratio $R(m_b)$ lower then threshold $\Theta$ are removed from blank in further computation. Let mark the new set of mass-to-charge ratio with max-to-mean ratio $R$ higher then threshold $\Theta$ as $\tilde{M}$:
\begin{equation}
\tilde{M_B} = M_B - \{m_b~:~ R(m_b) < \Theta\}, m_b \in M_B \label{eq:matilda}
\end{equation}
where $M_B$ is ordered set of $[m/z]$ values in the blank measurement, $R(m_b)$ is max-to-mean ratio (\cite{Chang, Bunting, Kohl}) and $\Theta$ is chosen threshold.
Videlicet, $\tilde{M_B}$ is just a subset of $M_B$ with property $R < \Theta$. However, the data reduction is not strictly necessary. Thresholding is not initial selection of alignment markers. It is just a simple random noise filtration. 

Also could be the ratio set $R$ separated only to lower and higher region by threshold equals to median value, whereas with threshold computed by equation (\ref{eq:th}) retain at least 2/3 of the blank measurement. In the blank with huge level of impurities may almost all data points pass through the thresholding, at least it still discards the low relevant of them (in meaning of capability for being markers in time-alignment).

\subsection*{Step 2.:Markers selection}
The second step is the foot-stone for all comparison tasks and it is known as the selection of the markers (\cite{LiX, Kirchner}). In other words, the markers are point candidates for the alignment itself. The markers in the approach are defined only from the blank, instead of searching for similar values in compared sets. Without any hesitation, it is sure that they are present in the sample measurement(s) also. Therefore, the corresponding data points can be easily pinpointed from the sample, after finished definition.

As was described above, in every measurement (even in the blank) is presented at least one of SP or WOT event. Successfully, SP occurs on the first half (in time axis) of the measurement and WOT on the second half (not considering peculiar operator errors like two measurements in one data set, stored only middle of measurement or nothing, etc.). Therefore, one can split the blank in time into two subparts (time intervals), each possibly containing one expressive feature. Using gradient changes during measurement offers splitting into more subparts (not necessary equidistant) with simple selection of cutting times. Just be sure, that the distinctive baseline inflex point (local minimum or maximum in intensity) is somewhere in the middle of selected interval (or leastwise not exactly on the interval borders). And one know the exact time value of that inflex point from the settings of the experiment, it was designed such. Past question, maximal number of time intervals is equals to the number of measured time points in the discrete data set, i.e. equals to the cardinality ($\aleph$) of set $T_B$. The optimal number of subparts could be determined by statistically appropriate methods (\cite{Perillo, Hyndman}), in case of equidistant intervals.
Let assume that sets $T_B$, fulfill the sampling theorem (\cite{Nyquist, Kotelnikov, Shannon1, Shannon2}) and split the blank time axis (and therefore whole blank measurement) into $n$ equidistant subparts, where $2 \leq n \leq \aleph(T_B)$. For simple illuminating example, is $n$ equals to 3.  Now one obtain three time intervals $T1_B$, $T2_B$ and $T3_B$ (or $T\vartheta_B, \vartheta = {1,...,n}$ shortly) as the subsets of $T_B$:
\begin{equation}
(T1_B \subset T_B) \wedge (T2_B \subset T_B) \wedge (T3_B \subset T_B),\label{eq:subset}
\end{equation}

\begin{equation}
T1_B \wedge T2_B \wedge T3_B = T_B.\label{eq:subsets}
\end{equation}

The intervals are defined with additional properties.

I.) The sets $T\vartheta_B$ are increasingly ordered sets.

II.) time interval $T1_B$ precede time interval $T_B2$ and time interval $T2_B$ precede time interval $T3_B$:

\begin{equation}
T1_B \prec T2_B \prec T3_B.\label{eq:preciding}
\end{equation}

III.) The cardinalities of the subsets are equal or approximately equal:

\begin{equation}
\aleph (T1_B) \approx \aleph (T2_B) \approx \aleph (T3_B), \label{eq:cardinal}
\end{equation}

\begin{equation}
\aleph (T1_B) + \aleph (T2_B) + \aleph (T3_B) = \aleph (T_B), \label{eq:sumcardinal}
\end{equation}

because the time intervals $T\vartheta_B$ are equidistant or semi-equidistant (if cardinality of $T_B$ is or is not divisible by $n=3$ in natural numbers $\mathbb{N}$). In the worst case, cardinality of the shortest time interval differs to the others only by one.

The most common and understandable representations of LC-MS measurements are Total Ion Chromatogram (TIC) and mass spectrum. Mass spectrum is a measure of MS detector signal (intensities $y$) versus mass-to-charge ratio axis ($m \in M$ or $\tilde{m} \in \tilde{M_B}$ in the example now). One mass spectrum is just a slice of selected time in the whole measurement. The amount of all individual mass spectra in the measurement is equal to the cardinality of the set $T$. Therefore, is also the amount of mass spectra in each time intervals $T\vartheta_B$ equals to the cardinality of the related interval. TIC is a measure of detector signal versus time axis $T_B$. It is amount of all intensity values $y$ in exact time point $t \in T_B$: 
\begin{equation} 
\gamma_B(t) = \sum_{\tilde{m}} {y(t,\tilde{m})}, y \in Y_B.  \label{eq:tic}
\end{equation}

So, one obtained three different sub-TICs $\gamma \vartheta_B$, after splitting the time axis $T_B$ into $n=3$ intervals:
\begin{equation} 
\gamma \vartheta_B (t\vartheta) = \sum_{\tilde{m}} {y(t\vartheta,\tilde{m})}, t\vartheta \in T\vartheta_B, \vartheta = {1,...,n},  \label{eq:ticnu}
\end{equation}
one blank sub-TIC $\gamma \vartheta_B$ for each time interval $T\vartheta_B$.

The splitting of the time set $T_B$ into $n$ subparts (time intervals) $T\vartheta_B$ and therefore splitting of TIC $\gamma_B$ into sub-TICs $\gamma \vartheta_B$ also define the amount of markers used for time-alignment. There is necessary only one point in each time interval and it is almost directly selected from the related blank sub-TIC. 
As a blank marker is considered the time value $\tau_B$ of the subset $T \vartheta_B$, where the sub-TIC value is the maximal value of that sub-TIC:
\begin{equation} 
\tau_B(\vartheta) \mid \gamma \vartheta_B (\tau_B(\vartheta)) = max( \gamma \vartheta_B (t \vartheta)), \tau_B(\vartheta) \in T\vartheta_B.  \label{eq:maxgamma}
\end{equation}
In other words, is in time point $\tau_B(\vartheta)$ significant inflex point of blank sub-TIC $\gamma \vartheta_B$. Equation (\ref{eq:maxgamma}) produces the set $\{\tau_B\}$ of cardinality $\aleph = n$ as the set of blank markers for transformation function. Blank time axis $T_B$ is in this approach considered as reference time axis for each time-alignment of measurement done with similar experiment conditions.

It is slightly trickier to identify corresponding markers in analyte measurement time axis $T_A$. The minimal and maximal values of measurement TIC $\gamma_A$:
\begin{equation} 
\gamma_A(t) = \sum_m {y(t,m)}, y \in Y_A,  \label{eq:ticM}
\end{equation}
occurred in different parts of measurement, because of presence of the analyte. Cardinality of measurement mass-to-charge ratio set $M_A$ is bigger then cardinality of blank mass-to-charge ratio set $M_B$. The reason is obvious, at least one $m_A$ value of the measured analyte was added into the mobile phase to make the experiment meaningful. Usually, the amount of added mass values is higher than one. There is not only the analyte molecular ion, but its isotopes, fragments molecule, adducts and impurities too. Therefore, cardinality of the intensities set $Y_A$ has to be also bigger than cardinality of set $Y_B$. Bigger amount of molecules with bigger amount of possible mass-to-charge ratios in almost same measurement time length ($T_A \approx T_B$) produce wider dynamic range of intensity set $Y_A$:
\begin{equation} 
\aleph(M_A) > \aleph(M_B) \wedge \aleph(Y_A) > \aleph(Y_B).  \label{eq:sets}
\end{equation}
Surprisingly, the analyte measurement TIC $\gamma_A$ is not relevant for selection of the analyte marker set $\{\tau_A\}$. The pinpointing process from sets $(T_A, M_A,Y_A)$ differs from blank.

One more set of information is necessary to extract from blank measurement. With the knowledge of when ($in~\tau_B(\vartheta)$) the maximal value of sub-TIC $\gamma \vartheta$ was obtained, is also profitable to ask where (in mass). Slice of selected time in the whole measurement (blank or analyte) represents the mass spectrum as tuple:
\begin{equation} 
y(t) = [y(t,m_{j})], m_{j} \in M, y \in Y.  \label{eq:spectrum}
\end{equation} 
Not every mass $m_{j}$ was presented in detector in selected spectrum, i.e. some of the intensity values $y(t,m_{j})$ are equal to zero in selected time. In mass spectrum is feasible that two different and distinguishable mass values reach the exactly same intensity ($y(t,m_q) = y(t,m_w),~q \neq w,~t = const.$). Equality in non zero intensity values is not very often, however there is nothing bizarre on this event. The probability is small, but it does not mean impossibility of the event, especially in huge amount of different molecules detected by MS during the measurement. Hence, the mass spectrum is described as tuple and not as a set.

In time markers $\{\tau_B\}$ are corresponding $n$ mass spectra tuples $y(\tau_B)$ of the blank. As a $\vartheta-th$ blank mass marker is considered the mass-to-charge value $\eta_B(\vartheta)$ of the set $M_B$, where in the mass spectrum  $y(\tau_B(\vartheta))$ is the maximal value of intensity:
\begin{equation} 
\eta_B(\vartheta) \mid y(\tau_B(\vartheta), \eta_B(\vartheta))= max([y(\tau_B(\vartheta),\tilde{m_b})]), \tilde{m_b} \in \tilde{M_B}, y \in Y_B.  \label{eq:mxmass}
\end{equation} 
The cardinalities of blank time and mass markers are equal:
\begin{equation} 
\aleph(\{\tau_B\}) = \aleph(\{\eta_B\}),  \label{eq:cardieq}
\end{equation} 
and time values $\tau_B(\vartheta)$ with mass values $\eta_B(\vartheta)$ make set of whole blank markers as $n$ ordered pairs $\{(\tau_B,\eta_B)\}$.

Analyte measurement time axis $T_A$ is also separated into $n$ intervals $T \vartheta_A$, $\vartheta = 1..n$. Each analyte interval is approximate (means very similar) to blank interval ($T\vartheta_A \approx T\vartheta_B$) in equidistant case with approximately same start and end time point of the measurement ($T_A \approx T_B$). It is necessary to carefully choose the individual interval borders, when the time splitting was based on gradient inflex points. Corresponding gradient changes have to be situated in corresponding time intervals. Correct separation task could be simplify by proper timing of all measurements recording process and equipment synchronization. 

Direction of analyte markers selection is opposite to the blank situation - from mass to time values. As analyte mass markers are considered blank mass-to-charge ratios $\{\eta_B\}$ that are present in the analyte mass set $M_A$:
\begin{equation} 
\eta_A(\vartheta) \mid \eta_A(\vartheta) = \eta_B(\vartheta), \label{eq:analmass}
\end{equation} 
\begin{equation} 
\eta_B(\vartheta) \in \tilde{M_B} \wedge \eta_B(\vartheta) \in M_A \Leftrightarrow \eta_A(\vartheta) \in M_A \wedge \eta_A(\vartheta) \in \tilde{M_B}. \label{eq:analmass2}
\end{equation} 

Mass-to-charge ratios $\{\eta_B\}$ are supposed to be in the analyte measurements set $M_A$. Values $\eta \vartheta_B$ were taken from the blank set $\tilde{M_B}$ and belong to the molecules of mobile phase. Mobile phase is a part and parcel of the analyte measurement. This condition is always fulfilled if whole blank markers selection was done on mass-to-charge subset $\tilde{\tilde{M_B}}$:
\begin{equation} 
\tilde{\tilde{M_B}} \subset \tilde{M_B} \subset M_B \mid \tilde{\tilde{M_B}} = \tilde{M_B} \cap M_A. \label{eq:intersect2}
\end{equation} 
In other words subset $\tilde{\tilde{M_B}}$ is defined as intersection of blank mass subset $\tilde{M_B}$ from Step1 and analyte mass set $M_A$. Therefore, values $\tilde{\tilde{m_b}}$ are present also in blank measurement and analyte measurement:
\begin{equation} 
\tilde{\tilde{m_b}} \in \tilde{\tilde{M_B}} \Leftrightarrow \tilde{\tilde{m_b}} \in \tilde{M_B} \Leftrightarrow \tilde{\tilde{m_b}} \in M_A. \label{eq:massboth}
\end{equation} 
Instead of $\tilde{M_B}$ or $\tilde{m_b}$ is used $\tilde{\tilde{M_B}}$ or $\tilde{\tilde{m_b}}$ respectively in equations (\ref{eq:tic}..\ref{eq:analmass2}). Thus, is redundant to distinguish signs $\eta_B$ and $\eta_A$, because both tuples are equal. Let sign mass markers for further purpose only as $\eta$:
\begin{equation} 
\eta(\vartheta) = \eta_B(\vartheta) = \eta_A(\vartheta) \mid \forall \vartheta=1..n \Rightarrow \{\eta\} = \{\eta_B\} = \{\eta_A\}. \label{eq:etas}
\end{equation} 

That is not as trivial as seems to be. Blank mass markers $\{\eta_B\}$ are values $\tilde{m_b}$ or $\tilde{\tilde{m_b}}$ from the subset $\tilde{M_B}$ or $\tilde{\tilde{M_B}}$ respectively. On the other hand, analyte mass markers $\{\eta_A\}$ are values from the set $M_A$. Therefore indexes $b$ and $a$ are not equal, even if the value $m_b$ equals to the value $m_a$. Obviously, there is forbidden the exception of special case where set $M_B$ or $\tilde{M_B}$ or $\tilde{\tilde{M_B}}$ respectively strictly equals to the set $M_A$, for two serious reasons. At first, set $M_A$ contains additional mass values of the analyte itself, not presented in blank measurement. At second, some random noise is always presented. The probability is extremely low in our universe, that two measurements have exactly the same distribution of random noise occurrence which fits in values and positions. Sign simplification done by equation (\ref{eq:etas}) is allowed just because blank mass subset $\tilde{\tilde{m_b}}$ is no more necessary in time-alignment process. However, $b$ and $a$ indexes inequality is important to consider in algorithm implementation (wrong index is one of the top common source code mistakes in programs development).

Only a part of analyte measurement is further investigated, once the mass markers $\{\eta\}$ were pinpointed. The behavior of single analyte mass value $m_a$ in time could be described as mapping from that mass value $m_a \in M_A$ and the set $T_A$ into the set $Y_A$ of intensity values $y$. This mapping process produce Single Ion Chromatogram (SIC) as a function of time:
\begin{equation} 
\gamma_{m_a} (t) = y(t,m_a), ~t \in T_A~,~y \in Y_A.  \label{eq:sic}
\end{equation}
Therefore, for each mass value $m_a$ from set $M_A$ exist one SIC ($\aleph (\{\gamma_{m_a}\}) = \aleph (M_A)$). Consequently, the analyte TIC $\gamma_A(t)$ is just a sum over $m_a \in M_A$ of all analyte SICs $\gamma_{m_a} (t)$:
\begin{equation} 
\gamma_A(t) = \sum_{m_a} \gamma_{m_a} (t) = \sum_{m_a} y(t,m_a), m_a \in M_A,~t \in T_A,~y \in Y_A.  \label{eq:Atic}
\end{equation}

Note, that it is seemingly skipped the step of analyte measurement points reduction. In case of mass markers $\eta \in M_A$ is necessary only $n$ number of analyte SICs, just $\gamma_{\eta(\vartheta)}$:
\begin{equation} 
\gamma_{\eta(\vartheta)} (t) = y(t,\eta(\vartheta)), t \in T_A,~y \in Y_A.  \label{eq:etasic}
\end{equation}

Therefore, decreasing of amount of points in analyte measurement is greater in contrast to the blank measurement reduction in Step 1. ($\aleph(\{\eta\}) \ll \aleph(M_A)$). Moreover, not whole SIC $\gamma_{\eta(\vartheta)}$ is required for selection of $\vartheta-th$ analyte time marker $\tau_A(\vartheta)$. The analyte measurement time axis $T_A$ was separated into $n$ intervals $T \vartheta_A$. It is quaranted to find the $\vartheta-th$ time value $\tau_A$ in time interval $T\vartheta_A$, when the time set separation was done correctly ($T\vartheta_A \approx T\vartheta_B$). Thus, analyte time markers pinpointing process works on $n$ sub-SICs, instead of whole analyte measurement ($(T_{A}, M_{A}, Y_{A})$). The $\vartheta-th$ sub-SIC is then defined as a part of mass marker $\eta(\vartheta)$ SIC $\gamma_{\eta(\vartheta)} (t)$ on time interval $T\vartheta_A$:
\begin{equation} 
\gamma \vartheta_{\eta(\vartheta)} (t\vartheta) = y(t\vartheta,\eta(\vartheta)), t\vartheta \in T\vartheta_A,~y \in Y_A,\vartheta = {1,...,n}.  \label{eq:subsic}
\end{equation}

\begin{figure*}[ht]
\centering
\includegraphics[scale=0.27]{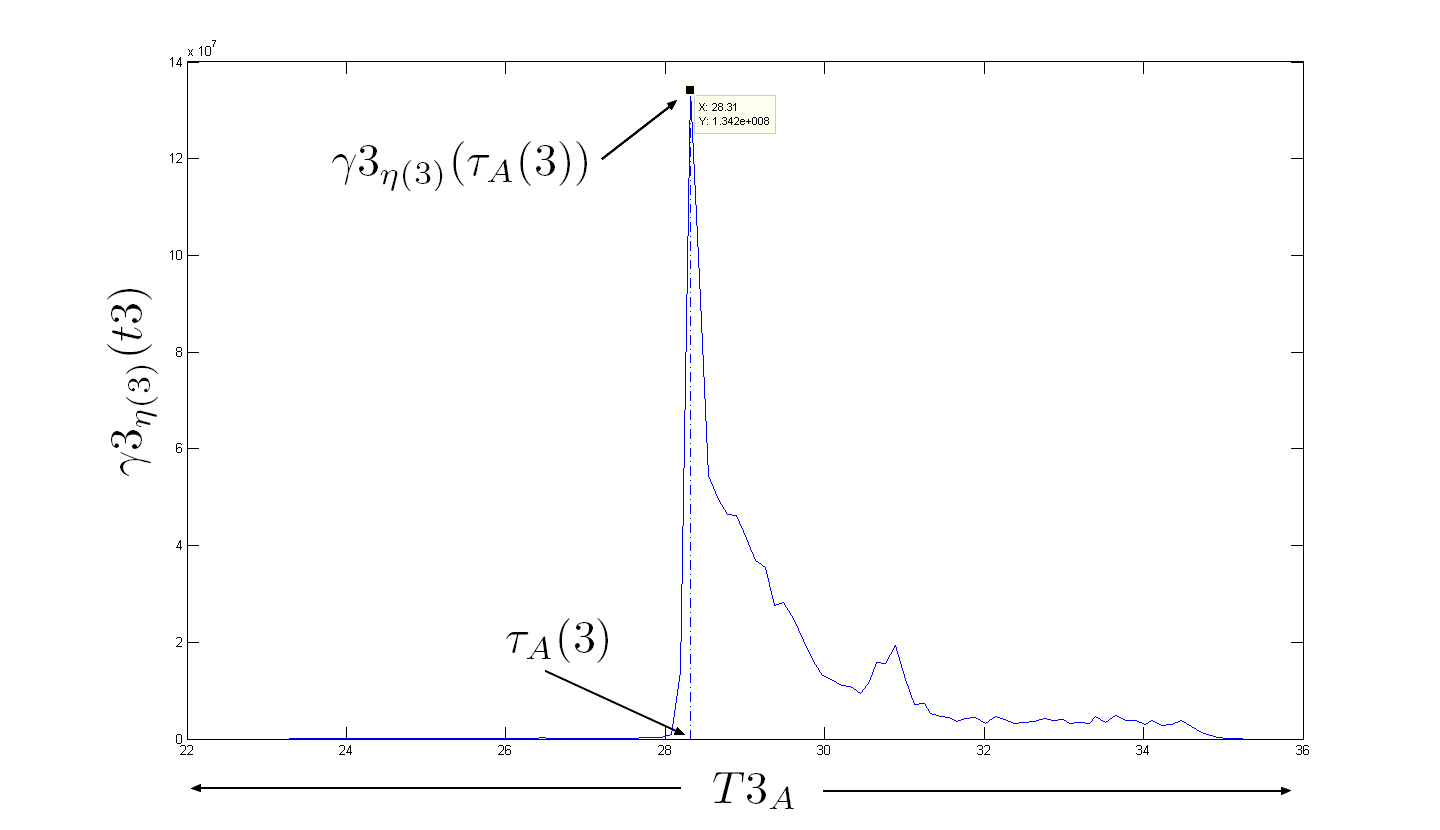}
\caption{Example of analyte time marker selection. In the $3-rd$ sub-SIC $\gamma3$ of the analyte mass $\eta(3)$ is maximal intensity obtained in the time value $\tau_A(3)$. Therefore, the $3-rd$ analyte time marker $\tau_A(3)$ value is equals to $28.31 ~[min]$ in this example. There is no mass spectrum, because SIC consist (by its definition) of single $[m/z]$ value $=~\eta(3)$.} 
\label{subsic.graphicx}
\end{figure*}

As an analyte time marker is considered the time value $\tau_A$ of the subset $T \vartheta_A$, where the sub-SIC value $\gamma \vartheta_{\eta(\vartheta)}$ is the maximal value of that sub-SIC:
\begin{equation} 
\tau_A(\vartheta) \mid \gamma \vartheta_{\eta(\vartheta)} (\tau_A(\vartheta)) = max( \gamma \vartheta_{\eta(\vartheta)} (t\vartheta)), \tau_A(\vartheta) \in T\vartheta_A.  \label{eq:maxsic}
\end{equation}

\begin{figure*}[hp]
\centering
\includegraphics[scale=0.20]{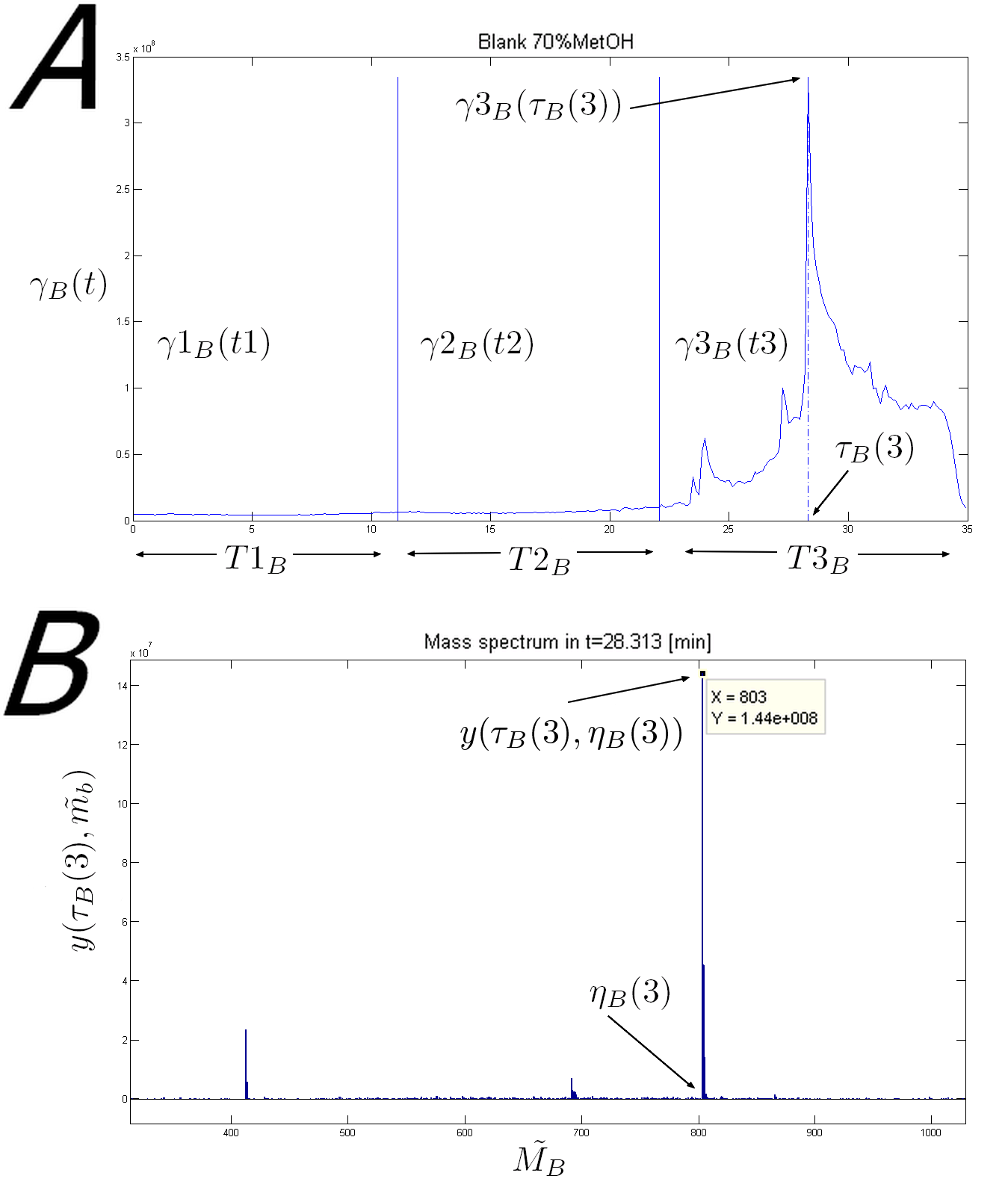}
\caption{Example of blank markers selection. Panel 2A shows Total Ion Chromatogram (TIC) $\gamma_B(t)$ separated into $n=3$ sub-TICs $\gamma 1_B$, $\gamma 2_B$ and $\gamma 3_B$ on time intervals $T1_B$, $T2_B$ and $T3_B$. Maximal intensity value $\gamma 3_B (\tau_B(3))$ is in time interval $T3_B$ located on time $\tau_B(3)$. Panel 2B shows mass spectrum in selected time $\tau_B(3)$. Maximal intensity $y(\tau_B(3),\eta_B(3))$ is obtained on mass $\eta_B(3) \in \tilde{M_B}$. Blank time marker value $\tau_B(3)$ is equals to $28.313~[min]$ and blank mass marker value $\eta_B(3)$ is equals to $803~[m/z]$ in this example. Apparently, there are no visible relevant features for markers selection. However, the range of intensity axis is $10^8$, which dissable details in lower intensity values. That is exactly why observation only of TICs is not wisdom.} 
\label{maxima.graphicx}
\end{figure*}

The total space of values to be analyzed is rapidly decreased (from thousands to ones). Process of the selection of the markers is indicated on Figure \ref{maxima.graphicx}, for mathematical details and justification see chapter Methods. This is sufficiently robust approach because all blanks have discernible signals, even a watter (at least injection peak, however there are useful changes in span on the time axis). Once again, the determination of markers is enough to be done in blank processing and then pinpoint the corresponding markers in the analyte measurements.

Again, the cardinalities of analyte time and mass markers are equal:
\begin{equation} 
\aleph(\{\tau_A\}) = \aleph(\{\eta\}).  \label{eq:cardi2eq}
\end{equation} 
and mass values $\eta(\vartheta)$ with time values $\tau_A(\vartheta)$ make set of whole analyte markers as $n$ ordered pairs $\{(\tau_A,\eta)\}$.
It follows from the equation (\ref{eq:etas}) that mass markers $\eta$ are the same for blank and analyte. Therefore, (using equations (\ref{eq:cardi2eq}) and (\ref{eq:cardieq})) is also the amount of blank time markers equal to the amount of analyte time markers:
\begin{equation} 
\aleph(\{\tau_A\}) = \aleph(\{\tau_B\}) = n.  \label{eq:taueq}
\end{equation} 
This is exactly what is often demand (to have the same cardinality of two corresponding time sets) and makes the next step as easy as possible.

\subsection*{Step 3.:Transformation function(s)}

Finally, the third step works with the time values of the selected markers from both sets (blank and sample), which are now of the same cardinality and in the same order. This last step actually produces the transformation function, it computes the description of the time-alignment. However, the procedure is not limited to the given algorithm. Nonlinear shifts in the retention time between measurements arise especially from stochastic changes in column chemistry over time and minor changes (also stochastic) in mobile phase composition (\cite{Johnson, Norton, Podwojski}). Considering this nonlinearity between time axes leads to the various normalization rules or shift corrections (\cite{Kirchner, LiX}). The blank measurement time axis $T_B$ is considered as the reference time axis, in this approach. Generally, any analyte measurement time axis could be aligned onto blank time axis by a$~$priori unknown non-linear transformation function $\mathcal{F}$:
\begin{equation} 
t_b = \mathcal{F} (t_a,\beta), t_b \in T_B, t_a \in T_A, \{\beta\} \in \mathbb{R}, \label{eq:timefunc}
\end{equation} 
where $\beta$ denotes unknown parameter(s) of the function $\mathcal{F}$.

There is no strictly restriction for analyte time axis to be also considered as the reference one. Consequently, the blank measurement time axis could be aligned onto analyte time axis as by function $\breve{\mathcal{F}}$:
\begin{equation} 
t_a = \breve{\mathcal{F}} (t_b,\breve{\beta}), t_a \in T_A, t_b \in T_B,  \{\breve{\beta}\} \in \mathbb{R}, \label{eq:timefunc2}
\end{equation} 
and sign $\breve{\beta}$ denotes unknown parameter(s) of $\breve{\mathcal{F}}$, analogously.
Function $\breve{\mathcal{F}}$ is in ideal case (in deterministic world without noise where all processes are purely equilibristic infinitesimal changes in non-fractal phase space) identical to the inverse function $\mathcal{F}^{-1}$ of $\mathcal{F}$. However, it may be misleading to select one of the analyte measurements time axis. There has to be very pertinent reason for using equation (\ref{eq:timefunc2}). Exempli gratia, using time axis of healthy patient blood sample as reference time axis for other 'sick' patients is just a wish for experiment purpose. The simplest standard is still represented by the blank for chosen setup of measurement device (LC column, solvents, gradient changes, MS ionization, detector focus, and so on). Once again, blank is general basic information independent on the experiment higher-level interpretation. Vice versa, the blank measurement depends only on the experiment setup and device properties. Therefore, correct and rigorous blank measurement $(T_B, M_B, Y_B)$ describes the experiment. It is the knowledge ready to be used in time-alignment.

The transformation $\mathcal{F}$ is a description for adjustment of time axes relation. Time markers $\tau_B \in T_B$ and $\tau_A \in T_A$ are time values with superb property - the resemblance between $\tau_B(\vartheta)$ and $\tau_A(\vartheta)$ is congruent:
\begin{equation} 
\tau_B(\vartheta) \cong \tau_A(\vartheta), ~\forall \vartheta = {1,...,n}. \label{eq:congru}
\end{equation} 
In other words, time markers $\tau_B(\vartheta)$ and $\tau_A(\vartheta)$ match together. For the sake of completeness, relation between blank time axis $T_B$ and analyte time axis $T_A$ is homomorphism (structure-preserving mapping) and relation between time markers $\{\tau_B\}$ and $\{\tau_A\}$ is isomorphism (bijective homomorphism).

The most puzzling issue is the task of function $\mathcal{F}$ type specification (\cite{Giatting, Nyholt, Forni}), i.e searching for data analysis process for constructing mathematical mapping, that minimizes displacement of the data points (time values). Common approach is to create a class of possible models, but it is not always obvious what models should be used (\cite{Zcychaluk}). Even with the understanding of underlying physical and chemical properties of the problem is difficult to choose the right model. Hence, both in linear and nonlinear modeling is used regression analysis (\cite{Polettini}) as investigation of the hypothesis about the relationship between the variables of interest. Specific cases are various iterative methods for value interpolation (\cite{Prince, Cannataro}), in which the function must go exactly through the time markers $\tau$. The objective of regression analysis is to produce an estimate of the hidden parameters $\beta$ (\cite{Sykes}). Unfortunately, any parameter analysis can only help in differentiating between hypothesis or models (\cite{Ledvij}). Very strong results still do not prove that the correct function $\mathcal{F}$ was chosen (\cite{Reed}).

Note, that the linear functions are just the evaluation of polynomial of first degree. Consequently, the very first 'non-linearization' is the polynomial of higher degree. Insofar that, the most extremely primitive nonlinear function evaluate polynomial of second degree. The collection of eventual type of relations (models, mappings, hypothesis, functions, whatever) is huge. Harmonic analysis (wavelets, fast Fourier transformation, eigenvalues) and MVA are the famous and prevalent theories nowadays (\cite{Gathen, Childs, Martens2, Martens1}).

Therefore, the task of the proper transformation function selection is always nontrivial. For instance, the mentioned simple function was chosen to illuminate the power of blank measurement. Accordingly, the relation between blank time set $T_B$ and analyte time set $T_A$ is considered as polynomial function of second degree:
\begin{equation} 
\mathcal{F}(t_a, \beta):~t_b = \beta_2 {t_a}^2 + \beta_1 t_a + \beta_0 + \varepsilon_a, t_b \in T_B, t_a \in T_A, \beta_{\kappa} \in \mathbb{R}, k = {0,...,2}, \label{eq:polynom}
\end{equation} 
where $\varepsilon_a \in \mathbb{R}$ is an unobserved random variable, representing the errors in the data.
Let define the parameters vector $[\beta]$, blank time markers vector $[\tau_B]$ and analyte time markers $[\tau_A]$ matrix:
\begin{displaymath}
\left[ \beta \right]=
\left(
\begin{array}{c}
\beta_p \\
\beta_{p-1} \\
\vdots \\
\beta_0
\end{array}
\right)
,~
\left[ \tau_B \right]=
\left(
\begin{array}{c}
\tau_B(1) \\
\vdots \\
\tau_B(n) 
\end{array}
\right)
,
\end{displaymath}
\newline
\begin{displaymath}
\left[ \tau_A \right]=
\left(
\begin{array}{ccccc}
{\tau_A}^p(1) & {\tau_A}^{p-1}(1) & \cdots & \tau_A(1) & 1 \\
\vdots & \vdots & \ddots & \vdots & \vdots \\
{\tau_A}^p(n) & {\tau_A}^{p-1}(n) & \cdots & \tau_A(n) & 1
\end{array}
\right).
\end{displaymath}
where $p$ is degree of the polynomial (and therefore natural number, $p \in \mathbb{N}$) and $n$ is cardinality $\aleph$ of time markers $\tau_A$ or $\tau_B$ ($\aleph \{\tau_A\} = \aleph\{\tau_B\}$). In the example are $p=2$ and $n=3$.

The unknown parameters $\beta$ of polynomial transformation function $\mathcal{F}$ could be then estimated by regression analysis (using equation \ref{eq:congru}):
\begin{equation} 
[\beta] \simeq [A] \backslash [B],\label{eq:leftdivm1}
\end{equation} 
where sign $\backslash$ is defined as matrix left division
\begin{equation} 
[A] \backslash [B] = [A]^{-1} * [B], \label{eq:leftdivm2}
\end{equation} 
because matrix multiplication is not commutative.

	The problem is with the error $\varepsilon_a$, that causes only asymptotic equality in matrix equation (\ref{eq:leftdivm1}) and leads to the inexactly specified system of simultaneous equations. The solutions is a particular estimation of the values of all parameters $\beta$ that simultaneously satisfies all of the equations. Regression analysis offers numerous parameter estimation methods (\cite{Martens1, Martens2}), that differ in computational burdens and robustness depended on the distribution of unobserved error $\varepsilon_a$. Frequently used method to solving systems of equations is approach of least squares (\cite{Wolberg, Moler}). It is a technique that minimize the Euclidean length of a vector $[\varepsilon]$, defined as:
\begin{equation} 
[\varepsilon] = [A] * [\beta] - [B], \label{eq:erorta}
\end{equation} 	

This last step actually produces the parameters of transformation function, it computes the description $\mathcal{F}$ of the time-alignment:
\begin{equation} 
\grave{t_a}  = \beta_2 {t_a}^2 + \beta_1 t_a + \beta_0, \label{eq:ef}
\end{equation}
where time values $\grave{t_a} \in \grave{T_A}$ are analyte measurement time values $t_a \in T_A$ asymptotically aligned to the blank measurement time values:
\begin{equation} 
t_b \simeq \grave{t_a}. \label{eq:align}
\end{equation}

Furthermore, blank approach allows to align the time axes of all analyte measurements $(T_{A\lambda}, M_{A\lambda}, Y_{A\lambda})$, $\lambda \in \mathbb{N}$, done on the same chromatographic column under same experiment conditions. Simply, two given analyte time axis $T_{A1}$ and $T_{A2}$ are independently normalized to the blank time axis $T_B$:
\begin{equation} 
\grave{t_{a1}}  = \beta_{2(A1)} {t_{a1}}^2 + \beta_{1(A1)} t_{a1} + \beta_{0(A1)}, ~t_{a1} \in T_{A1},\label{eq:anone}
\end{equation}
\begin{equation} 
\grave{t_{a2}}  = \beta_{2(A2)} {t_{a2}}^2 + \beta_{1(A2)} t_{a2} + \beta_{0(A2)}, ~t_{a2} \in T_{A2},\label{eq:antwo}
\end{equation}
where $\beta_{\kappa(A\lambda)}$ are the parameters of polynomial transformation function $\mathcal{F}_{\lambda}$ of each analyte time axis $T_{A\lambda}$. Normalized time values $\grave{t_{a\lambda}}$ are asymptotically aligned to the time values $t_b$, by analogy of equation (\ref{eq:align}):
\begin{equation} 
t_b \simeq \grave{t_{a1}} \wedge t_b \simeq \grave{t_{a2}}. \label{eq:alignvta}
\end{equation}
Therefore, also time values $\grave{t_{a1}}$ are aligned to the time values $\grave{t_{a2}}$. 
\begin{equation} 
\grave{t_{a1}} \simeq \grave{t_{a2}}. \label{eq:alignimp}
\end{equation}
However, equation (\ref{eq:alignimp}) simplify any comparison of given analyte measurement $(T_{A\lambda}, M_{A\lambda}, Y_{A\lambda})$ using the knowledge of blank measurement $(T_B, M_B, Y_B)$ and estimated parameters $\beta_{\kappa(A\lambda)}$ of functions $\mathcal{F}_{\lambda}$.

The last two steps are very similar with DTW or IS. With standards addition, it is essential to locate their positions in the measurement data sets as input for time transformation function. The localization is algorithmically the comparison task, which is in principle time consuming and noise affected procedure. Some (or at least approximate) parameters of IS are known. This a priori information decreases slightly the complexity of comparison techniques. DTW is more difficult - the number of corresponding points in measurements is a priori unknown, data sets are large, impurities may be clear in signal but differ in order. Therefore, some filtration and preprocessing computation is optional. Of course, DTW could be also applied on IS to produce robust results, in case that IS are sufficiently dominant signals. Unfortunately, the strong and quick solutions are still far from quick and daily use in the rush lab during experiment tunning. As is shown in this chapter, BBTA has to deal only with minimal amount of selected points which are readily available.

\newpage
\section*{Results}
\addcontentsline{toc}{chapter}{Results}
Two analyte measurements $A1$ and $A2$ are aligned using BBTA. Thisr approach is compared with Correlation Optimized Warping (\cite{Tomasi}), one of the well known warping algorithm (\cite{Chae}). Both experimental samples were prepared by mixing methanolic extract of the cyanobacterium Nostoc sp. with the antifungal drug Nystatin $C_{47}H_{75}NO_{17}$ (Duchefa Biochemie, cat. no.: 003042.03). Nystatin was added into measurement $A1$ in concentration $= 0.5 [mg/ml]$ as compound with known value of molecular ion $= 926 [m/z]$. Nystatin in different concentration $= 0.05 [mg/ml]$ was added into measurement $A2$. 

The samples were analyzed on HPLC-MS (ESI) Agilent (\cite{agilent}) 1100 Series LC/MSD Trap using C8 reverse phase column (Zorbax XBD C8, $4.6 \times 150 [mm]$, $5 [\mu m]$) eluted by MeOH / Water gradient with addition of $0.1\%$ formic acid. The ion trap mass spectrometer was optimized for ions with [m/z] ratio 900 in positive mode. The data acquisition and exports were performed using ChemStation Software (Agilent) under WindowsNT operating system. The data analysis outputs were obtained by Expertomica metabolite profiling software (\cite{Urban}) under Windows XP/Vista operating system.

The spray needle was at a potential of $4.5 [kV]$, and a nitrogen sheet gas flow of 20 (arbitrary units) was used to stabilize the spray. The counter electrode was a heated $(200[^{\circ}C])$ stainless-steel capillary held at a potential of $10 [V]$. The tube-lens offset was $20 [V]$, and the electron multiplier voltage was $-800 [V]$. Helium gas was introduced into the ion trap at a pressure of $1 [mTorr]$ to improve the trapping efficiency of the sample ions introduced into the ion trap. The background helium gas also served as the collision gas during the collision activation dissociation (CAD).

Blank measurement $B$ was obtained without presence of the analyte mixture (Nostoc extraction, Nystatin). Therefore, Nystatin addition is not considered as IS due to its absence in the blank measurement. Only the blank itself represents internal standards in presented approach. 
The elements of time sets $T_{A1}, T_{A1}$ and $T_B$ differ to each other as is shown on \ref{tbl:times}. The cardinalities of analyte measurements are equal ($\aleph(\{T_{A1}\}) = \aleph(\{T_{A2}\}) = 322)$, the cardinality of blank measurement is lower ($\aleph (\{T_{B}\} = 313)$).

\begin{table}[h,t,b]
\begin{tabular}{|c||c|c|c|c|c|c|c|c|}
\hline
~& 1 & 2 & 3 & ... & 312 & 313 & 314\\
\hline
$t_{a1}$ & 0.0030 & 0.0963 & 0.1891 & ... & 33.7272 &  33.8422 & 33.9575\\
\hline
$t_{a2}$ & 0.0042 & 0.1018 & 0.1952 & ... & 33.8274 & 33.9436 & 34.0589\\
\hline
$t_b$ & 0.0042 & 0.1444 & 0.2265 & ... & 31.8125 & 31.9277 & $\emptyset$\\
\hline
\end{tabular}
\caption{Values of blank and analytes time sets values.}
\label{tbl:times}
\end{table}

\begin{figure*}[h,t,b,p]
\centering
\includegraphics[scale=0.20]{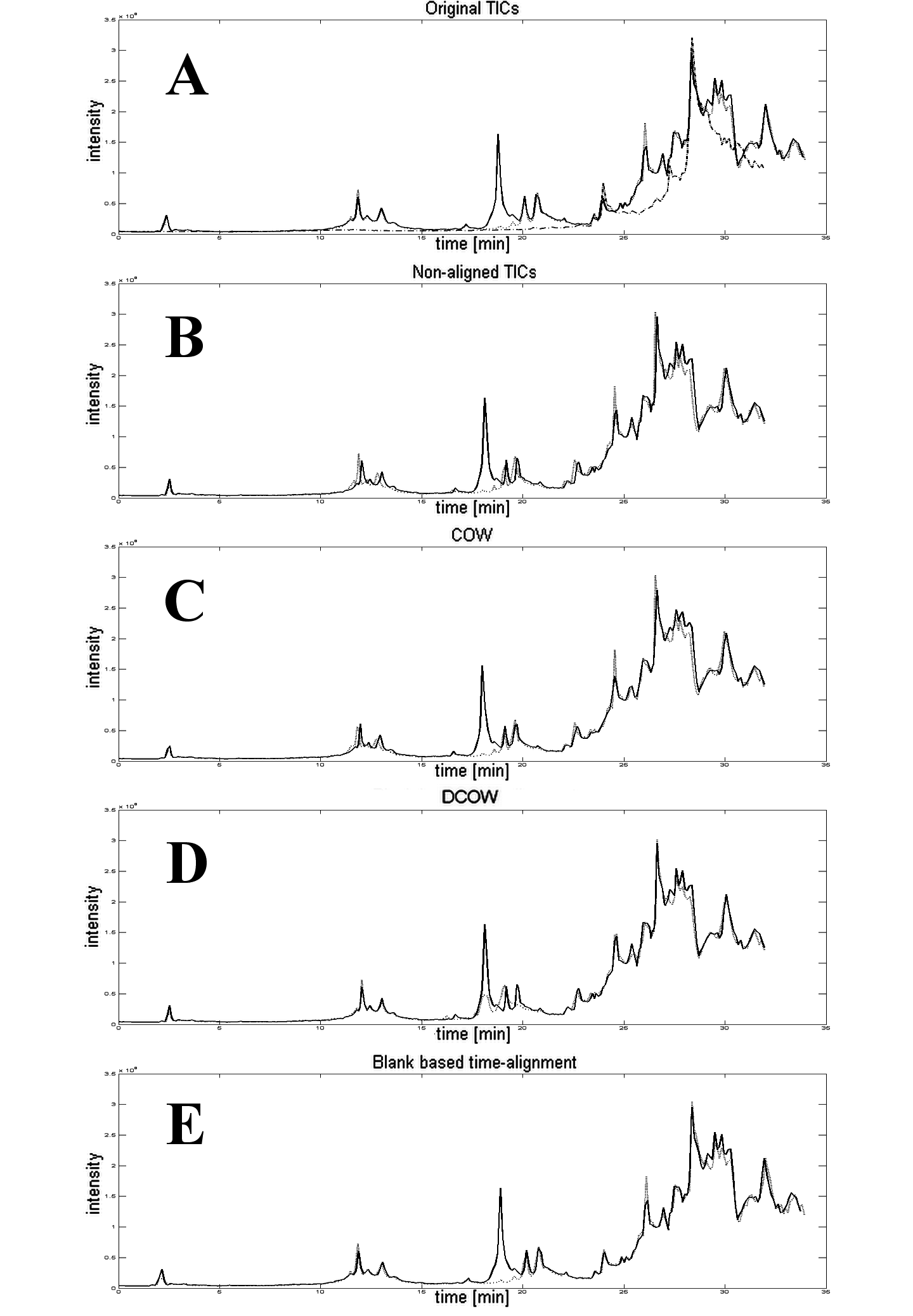}
\caption{Comparison of all TICs. Panel~A shows blank and analytes TICs $\gamma_B,\gamma_{A1},\gamma_{A2}$ in original time axes $T_B, T_{A1}, T_{A2}$. Panel~B shows artificially  dis-aligned analyte TICs $\gamma_{A1},\gamma_{A2}$ in reference time axis $T_R$. Panel~C shows results of analyte TICs $\gamma_{A1},\gamma_{A2}$ aligned to the blank TIC $\gamma_B$ by COW algorithm in reference time axis $T_R$. Panel~D shows results of analyt TIC $\gamma_{A2}$ aligned directly to the anlyte TIC $\gamma_{A1}$ by COW algorithm in reference time axis $T_R$. Panel~E shows results of analyte TICs $\gamma_{A1},\gamma_{A2}$ aligned to the reference time axis $T_R$ by Blank based time-alignment in aligned time axes $\grave{T_{A1}}, \grave{T_{A2}}$. Solid lines represents analyte TIC $\gamma_{A1}$, dotted lines represents analyte TIC $\gamma_{A2}$, dash-dotted line in panel~A represents blank TIC $\gamma_B$.}
\label{ta.graphicx}
\end{figure*}

The TICs of $A1$ (solid line), $A2$ (dotted line) and $B$ (dash-dotted line) are shown on \ref{ta.graphicx}A. Blank measurement $B$ is quite shorter by terminator of WOT decay beside to the analyte measurements $A1,A2$, as is clear from \ref{tbl:times} and \ref{ta.graphicx}A. Analyte measurements time axes were artificially dis-aligned by basic replacement to emphasize time shifts. In principle, analyte time axes are replaced by blank time axis. Let remind, that direct replacement has nothing to do with the alignment. Actually, it is the opposite process as is described further in this chapter.

Let denotes by sign $\varsigma$ the maximal amount of time elements in the given time sets: 
\begin{equation}
\varsigma = max(\aleph(\{T_{A1}\}),\aleph(\{T_{A2}\}),\aleph (\{T_{B}\})), \label{eq:maxcardinal}
\end{equation}
and slightly extend the definition of the reference time axis:
\begin{equation}
T_R  \mid T_B \subseteq T_R ~\wedge~ \aleph(\{T_R\}) = \varsigma. \label{eq:tref}
\end{equation}
The blank time $T_B$ is a subset of reference time set $T_R$ with cardinality equals to the $\varsigma$:

\begin{equation}
t_r \equiv t_b \mid r=b, t_r \in T_R, t_b \in T_B, r,b \in \{1,..., \aleph(\{T_B\}) \}. \label{eq:trisb}
\end{equation}

The missing time elements $\{t_{\aleph(\{T_B\})+1},...,t_{\varsigma}\} \in T_R$ could be set as equidistant continuation:
\begin{equation}
t_r = t_{\aleph(\{T_B\})} + \Delta t \times (r - \aleph(\{T_B\})), \label{eq:contin}
\end{equation}

where $\Delta t$ is estimated as averaging of difference between two consecutive time elements in blank time set $T_B$:
\begin{equation}
\Delta t = \frac{1}{\aleph(\{T_B\})-1} \sum_1^{\aleph(\{T_B\})-1} {(t_{\imath+1}-t_{\imath})}, ~t_{\imath+1},t_{\imath} \in T_B \label{eq:deltat}
\end{equation}

Theoretically, there are more easy ways how to create the reference time set $T_R$. Maximal operator in equation (\ref{eq:maxcardinal}) could be change into minimal and extension of equation (\ref{eq:contin}) is no longer necessary. However, minimal reference set means time data reduction and that is not advisable as it was in mass case (Step1. in Methods). The pinpointing process of the time markers $\tau$ is crucial part of time-alignment. Therefore, discarding time elements only for convenience reasons is dangerous way of thinking. No matter what the time elements values really are.
Another option, the addition at the beginning of the reference set $T_R$ is also possible, but complicated to no avail. The evaluation of missing time values and $\Delta t$ has the same computational burden (as addition at the end). However, the indexes $r$ has to be shifted and some of the added time elements may obtain negative values. The plots with negative time units on the reference time axis are not good exemplary candidates. The solution of setting all values added at the beginning to zero aims to the mismatch in TICs values. Therefore, is optional to follow the equations (\ref{eq:maxcardinal}...\ref{eq:deltat}).

Apparently, in the definition (\ref{eq:tref}) are missing some interval conditions. Time interval determined by minimal and maximal element of the reference time set $T_R$ should be congruently inside the time intervals determined by minimal and maximal elements of any given time sets. The truth of the matter is that in this example were the blank time set $T_B$ the set with minimal cardinality $\aleph({T_B}) < \varsigma$ and cardinalities of analyte measurements are both equal to the $\varsigma$. Furthermore, time interval congruent conditions are automatically fulfilled as is clear from the last row of \ref{tbl:times}. 

Equations (\ref{eq:maxcardinal}...\ref{eq:deltat}) as well as the reference time set $T_R$ are necessary just for the comparison of BBTA with COW, into the bargain. The purpose is to made this example and comparison as illustrative as possible. Hence, all values of analyte time elements $t_{a1}$ and $t_{a2}$ with indexes $a1$ and $a2$ in the range $<1..\varsigma>$ are replaced by the reference time values:
\begin{equation}
t_{a\lambda} := t_r \mid a\lambda = r, t_{a\lambda} \in T_{A\lambda}, t_r \in T_R, r \in \{1,...,\varsigma\}, \lambda = \{1,2\}. \label{eq:replace}
\end{equation}
Previous element values $t_{a\lambda}$ are forgotten. Description in equation (\ref{eq:replace}) produces \ref{tbl:reftimes}. All time sets $T_{A1}, T_{A2}, T_B$ and $T_R$ are now identical with also identical cardinality equals to $\varsigma$. However, the TIC values $\gamma_{A1}(t_r)$ and $\gamma_{A2}(t_r)$ corresponding to the $r$-th time element $t_r$ still differ to each other ($\gamma_{A1}(t_r) \neq \gamma_{A2}(t_r)$). The TICs did not change during time values replacing process:
\begin{equation}
\gamma_{A\lambda}(t_r) = \gamma_{A\lambda}(t_a) \mid r=a, t_r \in T_R, t_a \in T_{A\lambda}, ~\forall ~r,a \in \{1,...,\varsigma\}, \lambda = \{1,2\}. \label{eq:ticrep}
\end{equation}
Only the position of the TICs in the time axis has changed (\ref{ta.graphicx}B.). 

\begin{table}[h,t,b]
\begin{tabular}{|c||c|c|c|c|c|c|c|c|}
\hline
~& 1 & 2 & 3 & ... & 312 & 313 & 314\\
\hline
$t_{a1}$ & 0.0042 & 0.1444 & 0.2265 & ... & 31.8125 & 31.9277 & 32.030\\
\hline
$t_{a2}$ & 0.0042 & 0.1444 & 0.2265 & ... & 31.8125 & 31.9277 & 32.030\\
\hline
$t_b$ & 0.0042 & 0.1444 & 0.2265 & ... & 31.8125 & 31.9277 & 32.030\\
\hline
\end{tabular}
\caption{Time values of blank and analytes set to the reference time set.}
\label{tbl:reftimes}
\end{table}

\begin{figure*}[h,b,t]
\centering
\includegraphics[scale=0.3]{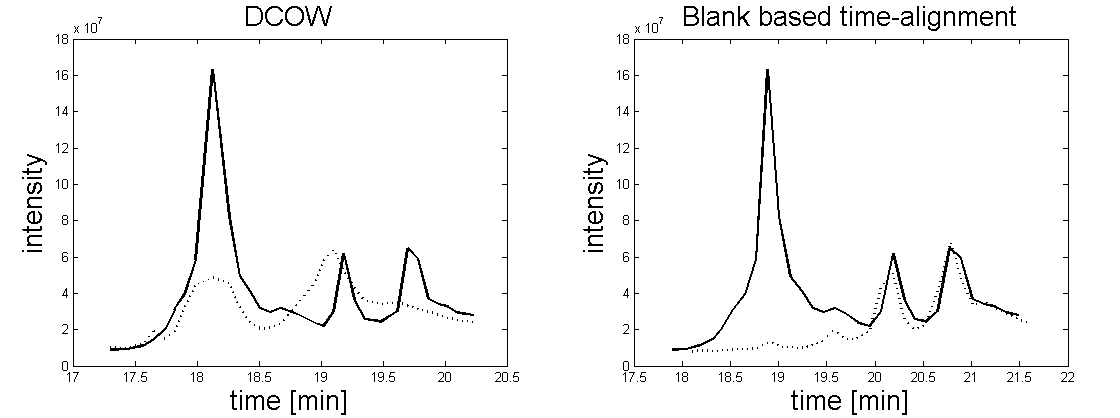}
\caption{Detail of Nystatin part of TICs  in DCOW  and BBTA. Analyte measurement $A2$ TIC (dotted line) was aligned to the analyte measurement $A1$ TIC (solid line). First of the two peaks after the Nystatin elution in $A2$ is incorrectly aligned to the Nystatin in $A1$ in DCOW.}
\label{nystatin.graphicx}
\end{figure*}

The COW algorithm aligns one or more data vector(s) onto reference vector via small changes in segments lengths on the data vector(s). Only the TICs values are considered as data vectors. For that reason, join reference time axis is required. Unfortunately, the time or mass sets are not taken into account in the available implementation (\cite{Tomasi}). Theoretical possibility of COW for all SICs in the measurements collides with input file limitation. There are over 2000 individual SICs in each measurements $B, A1, A2$. Two tunable parameters are necessary for COW, the number of segments (borders) and maximal increase or decrease of segment length (so-called slack). Optimal values of both parameters are estimated during the computation. The outputs of COW algorithm are aligned TICs $\grave{\gamma_{A1}}, \grave{\gamma_{A2}}$. 
Two variants of the COW algorithm were tested. The analyte measurements TICs $\gamma_{A1}, \gamma_{A2}$ were aligned to the blank TIC $\gamma_B$ in the first one (signed simply as COW). In the second one (signed as DCOW), the analyte measurement TIC $\gamma_{A2}$ was aligned directly to the analyte measurement TIC $\gamma_{A1}$.

The BBTA algorithm uses the three steps described in chapter Methods with default settings including automatic segmentation into three semi-equidistant segments and estimation of transformation function as polynomial function of second degree. Both analyte measurements TICs $\gamma_{A1}, \gamma_{A2}$ were aligned to the blank TIC $\gamma_B$ independently. The outputs are aligned time sets $\grave{T_{A1}}, \grave{T_{A2}}$. 

It is arduous to objectively evaluate the quality of any time-alignment. Comparison of the time values only is misguiding. The values are absolutely the same. Nevertheless, the corresponding TICs plots differ evidently. Another metric is so-called Peak integration error (\cite{Chae}) defined as:
\begin{equation}
PIE = abs(\frac{area_{aligned} - area_{non-aligned}}{area_{non-aligned}}) \times 100\% \label{eq:pie},
\end{equation}
where area is considered as integration of peak intensities. Therefore, area evaluation (and precision) is strictly dependent on used peak detection. Without any peak detector could be the area of whole measurement considered as input for equation (\ref{eq:pie}), for instance (\ref{tbl:compt}.). Blank based time-alignment changed only the time sets of the analyte measurements. There are no changes of the TICs values, no changes of the peaks (whatever they are), and no changes of the areas. For these reasons, the $PIE$ is nonsense in this case.

\begin{table}[h,t,b,p]
\begin{tabular}{|c||c|c|c|}
\hline
~ & COW & DCOW & BBTA\\
\hline
reference & $\gamma_B$ & $\gamma_{A1}$ & \scriptsize{$T_R$} \\
\hline
input data & $\gamma_B,\gamma_{A1},\gamma_{A2}$ & $\gamma_{A1},\gamma_{A2}$ & \scriptsize{$B,A1,A2$} \\
\hline
output data & $\grave{\gamma_{A1}}, \grave{\gamma_{A2}}$ & $\grave{\gamma_{A2}}$ & \scriptsize{$\grave{T_{A1}}, \grave{T_{A2}}$} \\
\hline
segments & 84 & 30 & 3 \\
\hline
slack & 1 & 13 & $\emptyset$ \\
\hline
time of computation & $\sim$ 3 [min] & $\sim$ 3 [min] & $\sim$ 140 [msec] \\
\hline
$PIE$ & 0.32\%& 0.67\%& 0.00\% \\
\hline
\end{tabular}  
\caption{Comparison of COW, DCOW and BBTA parameters. The main difference is in time of computation.}
\label{tbl:compt}
\end{table}

More objective metric of two similar LC-MS measurements is spectra comparison. A distance between a pair of spectra from two measurements in approximately same time has to be smaller in aligned case than in non-aligned one. Also the average distance between all spectra pairs (in corresponding time values) has to be smaller for aligned measurements. The only remaining question is the choice of distance evaluation method. It is beyond the scope of this work, to discuss the properties and pertinences of known distance metrics. The results of most common used formulas are shown in \ref{tbl:metrics}. In all cases are the spectra of BBTA closer together then in the non-aligned measurements. Naturally, optimal distance is equals to zero. However, the presence of random noise excludes the optimality in principal always.

\begin{table}[h,t,b,p]
\begin{tabular}{|c||c|c|c|c|c|c|c|}
\hline
~& eucl. & manh. & cos. & corr. & mink. & hamm. & cheb.\\
\hline
NA & $5.1\times10^6$ & $3.8\times10^7$ & 0.17 & 0.18 & $5.1\times10^6$ & 0.382 & $3.7\times10^6$\\
\hline
BBTA & $3.4\times10^6$ & $3.3\times10^7$ & 0.13 & 0.14 & $3.4\times10^6$ & 0.381 & $2.2\times10^6$\\
\hline
\end{tabular}\\
\caption{Average computed distance between pairs of spectra in non-aligned (NA) data and blank based time-aligned (BBTA) data.
Abbreviation: eucl.~-~Euclidean distance, manh~-~Manhattan distance (absolute difference), cos.~-~one minus angular cosine distance between spectra, corr.~-~one minus spectra linear correlation, mink.~-~Minkowski distance (generalization of both eucl. $\&$ manh. distance), hamm.~-~Hamming distance (\% values in spectra that are not identical), cheb.~-~Chebychev distance (maximal difference of values in spectra).}
\label{tbl:metrics}
\end{table}

\begin{figure*}[h,t,b,p]
\centering
\includegraphics[scale=0.16]{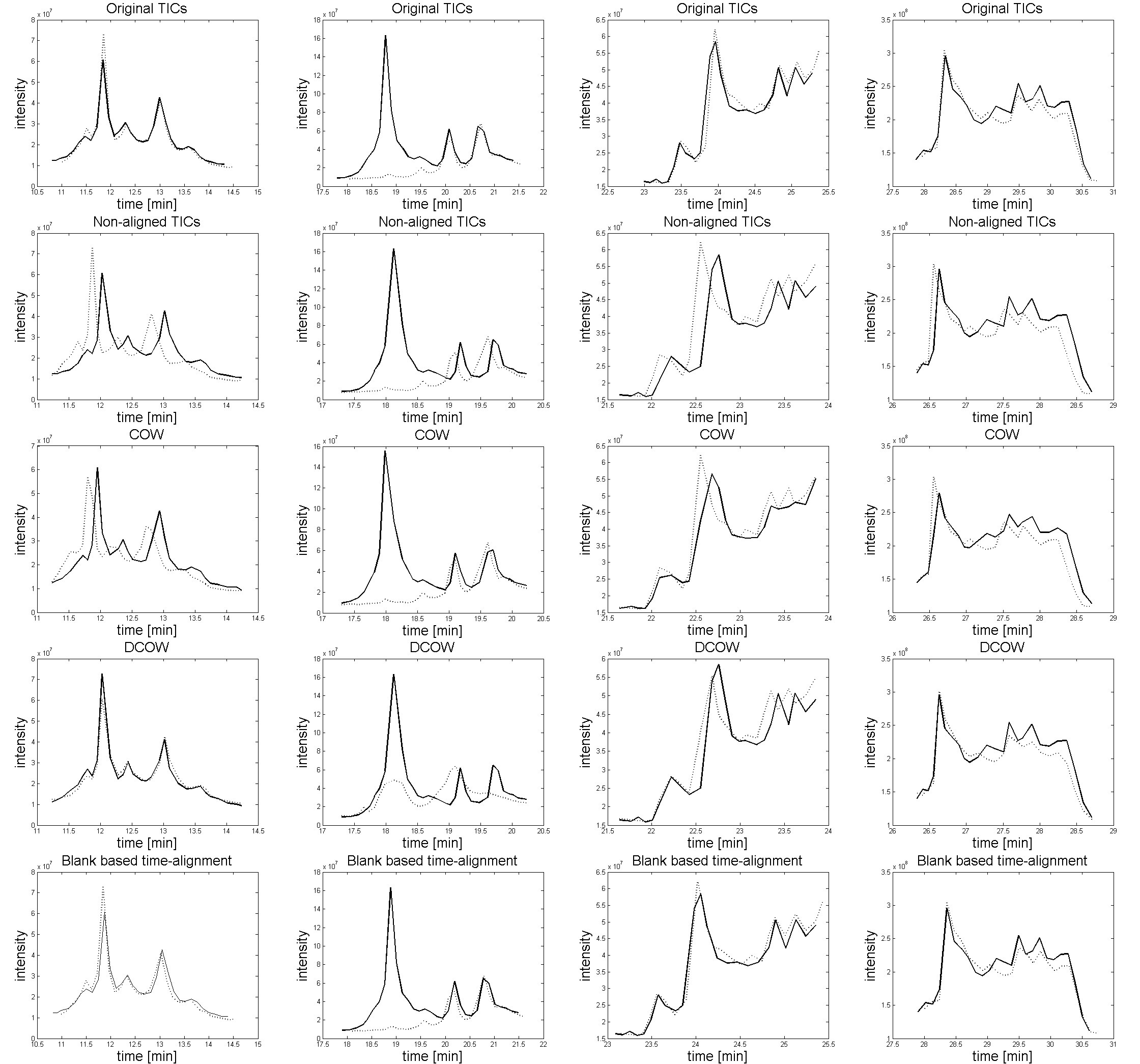}
\caption{Details of several TICs parts (columns). Rows from top to down: original TICs, non-aligned TICs, COW alignment, DCOW alignment and BBTA approach. The results of time alignments were computed on whole measurements. There are visualized only several parts of final plots to enhance differences between approaches.}
\label{details.graphicx}
\end{figure*}

Openly, the distinction between BBTA and COW alignment is quite unfair to the warping. The COW works only with the TICs, not with the whole measurements. However, full COW processing of all SICs exceeds the limits of available algorithm and may causes the mismatch in spectra. Obviously, the SICs can not be aligned to each other, the already pass together. The main problem with warps is more deeper and basic. Time warping is extremely powerful tool looking for parameters that minimize the distance between vectors. Therefore, it assumes that the alignment process is done for the same features that differ only in time duration and noise level. Thus, warp modification could be used as estimation for normalization function parameters as late as Step3, where the input warp features correspond to the time markers. Once again, using time warping directly on TICs confuses the algorithm unavoidably as it is shown on \ref{details.graphicx}. On the $2-nd$ column from the left, it is a part of TICs with Nystatin elution, which was described in Methods chapter. The concetration of Nystatin addition differs between analyte measurements $A1$ and $A2$. In COW case, there are analyte TICs aligned to the blank TIC. Therefore Nystatin can not affect the results in $3-rd$ row from the top of \ref{details.graphicx}. On the other hand, DCOW computes direct alignment of analyte measurement $A2$ TIC (dotted line) to the analyte measurement $A1$ TIC (solid line). As it is shown, one of the two peaks after the Nystatin elution in $A2$ is incorrectly aligned to the Nystatin in $A1$. That is not product of warping inefficiency, that is product of improper input. 

\begin{figure*}[h,t,b]
\centering
\includegraphics[scale=0.3]{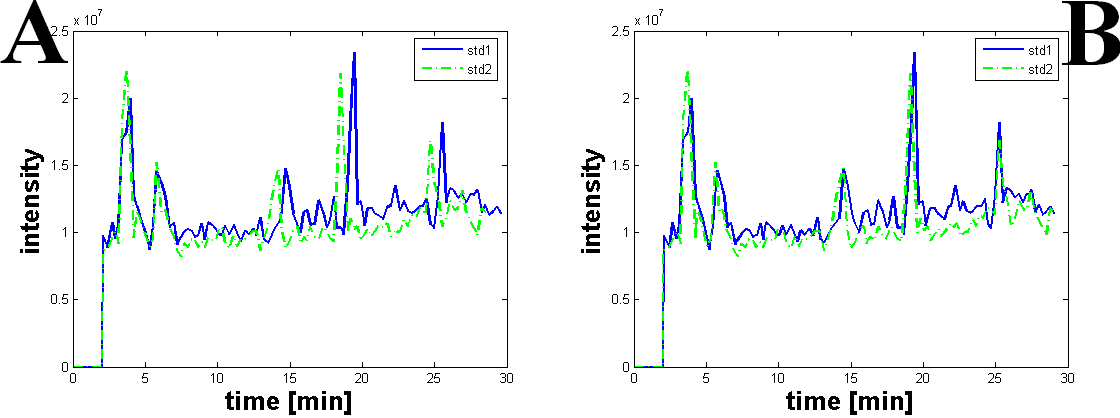}
\caption{Example of two mixture of standards ($std1$ and $std2$). As blank was used the same $H_2O$ as shown on Fig.\ref{blanks.graphicx}B without any standards addition. Panel A shows measurements before time alignment, panel B shows measurements after BBTA. Both measurements were aligned only to the blank, therefore there was no computation between $std1$ and $std2$.}
\label{std.graphicx}
\end{figure*}

It is necessary to emphasize the information that the BBTA approach works not only with the TICs. All markers selection process take into account whole measurement, therefore 3D matrix in time, mass and intensity space. It is also important, that markers selected from blank measurement are not usually significant in analyte measurement TIC, however they are still present in the matrix data. The BBTA approach is powerful enough to align data with simple blanks (with no patterns like peaks) even when the blank is just water (with some a priori unknown impurities) as is shown on \ref{std.graphicx}.

In comparison to the advantages of known time alignment methods the BBTA is also opened for extensions. Using blank as internal standard set is not in violation of additional standards. The blank measurement (and therefore the analyte measurements) could easily include addition of compounds estimated by LSERs (\cite{LiJ}). The markers $\tau$ pinpointed as relevant inflex points from blank in Step2 are just an optional subset of all eventual markers. For example, robust point matching known as Amsrpm (\cite{Kirchner}) is similar to the point of view to the systematic description of the measurements. Finally, exact analytical and parametric model for transformation function is complicated to define. In the example, in Step3 it is used polynomial of second degree. This primitive function demonstrate the power of blank based time alignment approach in comparison of COW. That was the key idea of this work. However, mathematically expressed, the space of function is unlimited as well as criterion evaluation. One of the semi-supervised warps is implemented in ChromA (\cite{Hoffman}). Unfortunately, ChromA is mainly focused on last step of time alignment. 
The BBTA premise measurements obtained by the same settings and devices. Thus, it is recommend to use geometric approach (\cite{Lange}) for comparison of different measurements from different devices.

In summary, it was used one of the most primitive normalization function for Step3 in simple example. Even then, the blank based time alignment results still prove blank usability. Step1 is not crucial for the approach, it is just for reduce of total time consumption. The main idea is presented in Step2. Selection of time markers with equal cardinalities solves problems with presumption fulfilling. Step3 is only regression analysis question and any algorithm belonging there could be improved. The idea of using blank measurements as internal standard is the main objective - the most simple and direct method for time alignment.

In contrast, all methods using peak detection for time alignment are error propagating (any error from the peak detection process is propagated into the further processing, obscure initial errors may emphasize errors in the output) \cite{Lindberg}. There, the time alignment strictly depends on ability of correct peak definition and detection. Possibly, that brings new set of dangerous presumptions into account. For example, in XCMS (\cite{XCMS}) toolbox for R are also used information from blank signals for time alignment. The ability of XCMS time alignment again depends on initially having a matching of peaks into reasonable groups (\cite{XCMS}). Moreover, XCMS approach of filtration change the shape of the peak according to the idealized model. Another example, a pre-processing tool for PARAFAC modeling (\cite{PARAFAC}) slightly extend the COW algorithm by correct idea of using covariance instead of correlation. The piecewise alignment similar to the COW was introduced by \cite{Pierce} with over-combined feature selection. However, warps might be easily confused by single metabolite, as it was shown in this chapter. Exhausting overview of both, commercial and freely available softwares for metabolomic data processing as well as time alignment was done by \cite{Katajama}. Some level of peak detection or binning is assumed in most of the available products.

Over and above, IS in sufficient amount will also fulfill this approach. Additional standards in the blank measurement constitute highly significant markers, if they were distinguishable by the column. However, IS addition is just the extension of BBTA. Basically, it is not necessary for the time alignment itself. The common usage is the support for identification. And that is certainly different problem.

All analysis computations were performed in Matlab (\cite{matlab}) 2008b on Intel CPU Centrino 2 P8600, 2.4 GHz, 4GB RAM.

\newpage
\section*{Conclusion}
\addcontentsline{toc}{chapter}{Conclusion}
BBTA is not general for comparison of any two or more measurements, but it is sufficient for measurements from the same chromatographic column with the same gradient settings. Nevertheless, these types of measurements represent everyday laboratory experiments in omics science, petroleum chemistry or pharmacology. One can directly afford the blank based approach, because of simple presumption. The mass values from the blank measurement are also presented in analyte measurement (or it can easily warrant it). Moreover, the time behaviors of the blank mass values are preserved in analyte measurements by the utilized settings. Hypothetically, if some corresponding time inflex point in the measurement was caused by the analyte mass, then the experiment was designed wrongfully. This situation can happen only when the blank mixture contains a compound with identical mass value to the analyte (but with different elution time).

The aspect of transformation function selection requires more consistent theory. However, it is a question of slightly different brand, especially nonlinear fits, regression analysis or genetic algorithms. This contribution still focused mainly on mechanism of simple, fast and reasonable markers definition from the blank measurement. 

Theoretically, this approach may also help to deal with the column aging. Mathematically, it is the problem of estimation of transformation between two or more blanks. When one of them is selected as the reference one, all other steps follow the described methods. Therefore, all analyte measurements could be aligned to the corresponding blank and hereupon aligned to the reference blank time axis. Unfortunately, data collection for column aging will take at least several months for everyday used column and years for rarely used column. 

BBTA is a mathematically derived and algorithmically simple approach for time alignment of 2D LC-MS chromatograms which requires blank measurement data. The principle is more objective than many methods known to us, inexpensive and readily available in any measurement series using the same procedure and devices. Moreover, all measurement spectra are preserved. Exemplificative transformation function could be easily supersede by any advanced estimation.

\end{document}